# Towards a complete, continuous, Wigner function for an ensemble of spins or qubits

Derek Harland[*]
*Department of Mathematics, Loughborough University,
Loughborough, Leics LE11 3TU, United Kingdom*

M.J. Everitt[†]
*Department of Physics, Loughborough University,
Loughborough, Leics LE11 3TU, United Kingdom*

Kae Nemoto and T. Tilma[‡]
*National Institute of Informatics, 2-1-2 Hitotsubashi, Chiyoda-ku, Tokyo 101-8430, Japan*

T.P. Spiller
*Quantum Information Science, School of Physics and Astronomy,
University of Leeds, Leeds LS2 9JT, United Kingdom*
(Dated: February 28, 2013)

We present a new quasi-probability distribution function for ensembles of spin-half particles or qubits that has many properties in common with Wigner's original function for systems of continuous variables. We show that this function provides clear and intuitive graphical representation of a wide variety of states, including Fock states, spin-coherent states, squeezed states, superpositions and statistical mixtures. Unlike previous attempts to represent ensembles of spins/qubits, this distribution is capable of simultaneously representing several angular momentum shells.

PACS numbers: 03.65.-w,03.65.Wj,03.65.Aa,02.30.Gp,02.20.-a,03.65.Ud

## I. INTRODUCTION

The drive to realise quantum technologies has led to great advances in experimental capacity. Techniques such as quantum state tomography are rapidly becoming just as much part of the experimentalists toolbox as the theorists. One of the most useful examples of such a tomographic method is the phase space distribution of the Wigner function [1]. One key element in this function's success is that it can be clearly linked to classical statistical mechanics and solutions to the Liouville equation, rendering in a clear and graphical manner the differences between classical and quantum states of a system. Indeed the Wigner function has been used to great effect in demonstrating the realisation of non-classical states of trapped light [2]. The usual application of the Wigner function is to systems that are expressed in terms of continuous variables. However, there has been some significant progress for finite dimensional systems where spin analogues of Wigner functions have been used to demonstrate entanglement in atom-chip circuits [3]. The usual approach here is to construct a pseudo probability distribution that maps information about the state of the system onto the surface of a sphere [4–9]. Such approaches have been very successful. For example, in [4] the authors provide a method for constructing Wigner-like functions on the sphere's surface producing intuitively useful pictures for Fock states and spin coherent states. This function can also be used to clearly demonstrate the non-classical nature of distinct superpositions of spin coherent states (so called spin-Schrödinger cat states) [10].

The Wigner function for systems of continuous variables is unique up to an arbitrary overall phase factor. Moreover, the Wigner function satisfies a number of other very important properties, for example its marginal distributions obtained through integration over a coordinate (e.g. position) provide the probability density in the canonically conjugate variable (e.g. momentum) [11–13]. While the previously mentioned spin Wigner-like functions are impressively good at providing intuitively meaningful pictures they do not satisfy all the properties of a true Wigner function. For example, distribution associated to the singlet state for two spins in [4, 5] is everywhere zero. Indeed these methods tend to only be good at analysing one total angular momentum subspace at a time (usually the largest one with, for $N$ spin-half particles, the angular momentum quantum number $l = N/2$). Hence, these phase space representations are not complete.

Such shortcomings are known and there have been various attempts to improve on these functions. These improved Wigner functions are usually defined on a discrete lattices or in higher dimensional phase spaces and tend to have most or all of the desired properties of Wigner functions [14–19]. However, there exists a *quid pro quo* and realising the full spectrum of properties of a Wigner function is done at the expense of not providing phys-ically intuitive pictures. While mathematically of real value, in our view, it is also very desirable to produce a function where there are obvious parallels between the quantum state and its graphical representation as is the case for Wigner's original function [1].

---
[*]Electronic address: d.g.harland@lboro.ac.uk
[†]Electronic address: m.j.everitt@lboro.ac.uk
[‡]Now at: Hiroo Gakuen, 5-1-14 Minami Azabu, Minato-ku, Tokyo, 106-0047, Japan

In this paper we partially address this problem by introducing a new, easy to compute, spin Wigner-like function that overcomes some of the difficulties found in [4–8] (such as being able to represent the singlet state for two spins). While not realising a complete representation of the state (and thus not being a true Wigner function), our function has many desirable features: it produces physically intuitive graphical representation; it can reproduce to good approximation previous Wigner-like distributions; and it is useful for a larger subspace of states than previous methods.

## II. STRATEGY

Our strategy for addressing this problem comprises several stages. First note that for $N$ spins $S^2$ has eigenvalues $l(l+1)$ given in terms of angular momentum quantum numbers $l = \frac{N}{2}, \frac{N}{2} - 1, \ldots$, such that $l \geq 0$. We also have the operator $S_z$ with quantum numbers $m = -l, -l+1, \ldots, l-1, l$. As $S^2$ and $S_z$ commute they share common eigenvectors that are indexed by these quantum numbers. It is a textbook result that for $N > 2$ the the quantum numbers (or, equivalently, eigenvalues) of $S^2$ and $S_z$ are not sufficient to uniquely determine a basis. This is due to the fact that $\{S^2, S_z\}$ is not a complete set of commuting observables (CSCO). We can, however, form a basis that can be uniquely labeled using the eigenvalues of a CSCO by introducing sufficient other observables with eigenvalue(s) $\mathbf{k}$ which commute with $S_x, S_y, S_z$. The vectors $\{|\mathbf{k}, l, m\rangle\}$ will then uniquely define an orthonormal basis (up to some arbitrary phase factor). However, when considering general properties of angular momentum it is common to use the shorthand $\{|l, m\rangle\}$ leaving $\mathbf{k}$ to be specified later. We will adopt this convention here.

Now we note that we can identify the two indices $l$ and $m$ with indices for two uncoupled harmonic oscillator Fock states $|n_1\rangle$ and $|n_2\rangle$. The idea being that, as it is straightforward to define Wigner functions for two harmonic oscillators, we can map the state of our system onto this Wigner function. However, because $\{S^2, S_z\}$ do not form a CSCO the labels $l$ and $m$ are insufficient to determine a unique mapping. This ambiguity may be removed by specifying, for each $l$, a particular value $\mathbf{k}_l$ for the eigenvalues (or quantum numbers) $\mathbf{k}$ of the additional observables needed to form a CSCO. A map $\Omega$ from the $N$-spin Hilbert space to the Hilbert space for coupled harmonic oscillators is then defined by

$$\Omega |\mathbf{k}, l, m\rangle = \begin{cases} |n_1 n_2\rangle = |l+m \ \ l-m\rangle & \mathbf{k} = \mathbf{k}_l \\ 0 & \mathbf{k} \neq \mathbf{k}_l. \end{cases} \quad (1)$$

The Wigner function $W(q_1, p_1, q_2, p_2)$ for a density matrix $\rho$ in its $l, m$ representation is then

$$\sum_{n_1', n_2', n_1, n_2} \langle n_1' n_2' | \Omega \rho \Omega^\dagger | n_1 n_2 \rangle W_{n_1 n_1'}(q_1, p_1) W_{n_2 n_2'}(q_2, p_2),$$

where $W_{nn'}(q, p)$ are Moyal functions [20–22], which will be discussed in more detail later on in the text.

While it may initially appear problematic that this Wigner function is four-dimensional it will turn out that, for many examples, we can make use of symmetry to reduce this Wigner function to a three dimensional distribution that will well suit our needs.

We note that $W(q_1, p_1, q_2, p_2)$ is defined in terms of the density operator. In principle we are able to replace this with any operator. Whether or not it is possible that, as with the Weyl-Wigner formulation, an equivalent description of quantum mechanics can be given entirely in terms of an extension to our Wigner function remains an open question. What is certain is that this can be done for systems comprising one or two spins.

## III. CONNECTION TO SPIN REPRESENTATION

We can define collective spin operators for a space of $N$ spins in terms of the Pauli operators, $\sigma_x$, $\sigma_y$ and $\sigma_z$ by

$$S_i = \frac{1}{2} \bigoplus_{k=1}^{N} \sigma_i^k \quad (2)$$

where $i = x, y$ or $z$ [34] and the superscript $k$ denotes the space to which the spins belong. These satisfy the $\mathfrak{su}(2)$ commutation relations,

$$[S_i, S_j] = i\epsilon_{ijk} S_k. \quad (3)$$

An eigenstate of $(S^2, S_3)$ with eigenvalues $(l(l+1), m)$ will be denoted $|lm\rangle$ (but note that for $l < N/2$ there may be many such eigenstates, more properly denoted $|\mathbf{k}, l, m\rangle$). We can now make use of Eq. (1) to map these eigenstates onto those of two harmonic oscillators and generate the associated Wigner function $W(q_1, p_1, q_2, p_2)$.

## IV. IMPLEMENTATION

As stated in section II our approach to constructing Wigner functions is to relate the system of spins to the two-dimensional quantum harmonic oscillator. Recall that a basis $|n_1 n_2\rangle$ for the Hilbert space can be constructed from the ground state $|00\rangle$ using ladder operators:

$$|n_1 n_2\rangle := \frac{1}{\sqrt{n_1! n_2!}} (a_1^\dagger)^{n_1} (a_2^\dagger)^{n_2} |00\rangle. \quad (4)$$

The ladder operators satisfy the relations,

$$[a_\alpha, a_\beta^\dagger] = \delta_{\alpha\beta}, \quad [a_\alpha, a_\beta] = 0, \quad [a_\alpha^\dagger, a_\beta^\dagger] = 0. \quad (5)$$

We will construct some new operators $J_1, J_2, J_3$ using the Pauli matrices $\sigma_i$ as follows:

$$J_i = \frac{1}{2} \begin{pmatrix} a_1^\dagger & a_2^\dagger \end{pmatrix} \sigma_i \begin{pmatrix} a_1 \\ a_2 \end{pmatrix}. \quad (6)$$

This may be thought of as an operator analogue of the Hopf fibration [23, 24] and is called the Jordan-Schwinger map [25, 26]. The reason for using this map to define $J_i$ is that they will then satisfy the same commutation relations as the $S_i$:

$$[J_i, J_j] = i\epsilon_{ijk} J_k. \quad (7)$$

It follows from Eq. (6) that the basis states $|n_1 n_2\rangle$ are also eigenstates for $J_3$ and $J^2 = J_1^2 + J_2^2 + J_3^2$:

$$J_3 |n_1 n_2\rangle = \left[\frac{n_1 - n_2}{2}\right] |n_1 n_2\rangle \quad (8)$$

$$J^2 |n_1 n_2\rangle = \left[\frac{n_1 + n_2}{2}\right] \left(\left[\frac{n_1 + n_2}{2}\right] + 1\right) |n_1 n_2\rangle. \quad (9)$$

In other words, the state $|n_1 n_2\rangle$ has angular momentum eigenvalues $l = (n_1 + n_2)/2$ and $m = (n_1 - n_2)/2$.

Thus we can now define a homomorphism by identifying a spin state's $l$ and $m$ eigenvalues with $n_1$ and $n_2$ according to the above prescription. This homomorphism maps the $N$-spin Hilbert space to the Hilbert space for the 2D harmonic oscillator and identifies the actions of $S_i$ and $J_i$, in particular sending eigenstates of $S^2, S_3$ to eigenstates of $J^2, J_3$.

The Wigner function for any state in the $N$-spin Hilbert space will by definition be the Wigner function of the associated state in the harmonic oscillator Hilbert space. The Wigner functions obtained will be 4-dimensional. We will show how this might work in practice in the next section by considering some examples.

For later use we briefly recall the Wigner functions for the harmonic oscillator [13]. The Wigner function for any system in one dimension with density operator $\rho$ is defined by the formula

$$W_\rho(q, p) = \frac{1}{\pi} \int dy \, \langle q - y | \rho | q + y \rangle e^{2ipy}. \quad (10)$$

In the particular case where $\rho = |n'\rangle\langle n|$ where $|n\rangle$ are eigenstates of the Hamiltonian for the 1D harmonic oscillator one finds

$$W_{nn'}(q, p) = \begin{cases} \frac{(-1)^n}{\pi} \sqrt{\frac{2^{n'} n!}{2^n n'!}} (q - ip)^{n'-n} e^{-(q^2+p^2)} L_n^{n'-n}(2(q^2+p^2)) & n \leq n' \\ \frac{(-1)^{n'}}{\pi} \sqrt{\frac{2^n n'!}{2^{n'} n!}} (q + ip)^{n-n'} e^{-(q^2+p^2)} L_{n'}^{n-n'}(2(q^2+p^2)) & n \geq n' \end{cases}, \quad (11)$$

with $L_n^\alpha$ denoting the Laguerre polynomials [27]. These are the Moyal functions for the harmonic oscillator [20–22]. The Wigner function in two dimensions is defined similarly:

$$W_\rho(q_1, p_1, q_2, p_2) = \frac{1}{\pi^2} \int dy_1 dy_2 \varrho(y_1, y_2) e^{2i(p_1 y_1 + p_2 y_2)}. \quad (12)$$

where $\varrho(y_1, y_2) = \langle q_1 - y_1, q_2 - y_2 | \rho | q_1 + y_1, q_2 + y_2 \rangle$. In the particular case where $\rho = |n_1' n_2'\rangle \langle n_1 n_2|$,

$$W_{n_1 n_2, n_1' n_2'}(q_1, p_1, q_2, p_2) = W_{n_1 n_1'}(q_1, p_1) W_{n_2 n_2'}(q_2, p_2). \quad (13)$$

We note that in [19] use was also made of two harmonic oscillators to map $|j = (n_1+n_2)/2, m = (n_1-n_2)/2\rangle$ to $|n_1 = j+m\rangle |n_2 = j-m\rangle$, however our focus is different as these authors then go on to relate their work to that of [4, 5, 7, 9].

## V. EXAMPLES

We now construct linear maps $\Omega$ from $N$-spin Hilbert spaces to the 2D harmonic oscillator Hilbert space, and use these to construct Wigner functions. We require that the maps $\Omega$ satisfy two conditions:

(i) $\Omega S_i |\psi\rangle = J_i \Omega |\psi\rangle$

(ii) $\Omega^\dagger \Omega$ is a projection.

Constructing $\Omega$ in this manner is equivalent to constructing $\Omega$ by introducing additional observables $\mathbf{k}$, as described in section II: condition (i) is equivalent to demanding that the additional observables commute with $S_i$; and condition (ii) is equivalent to assuming that the states $|\mathbf{k}, l, m\rangle$ are normalised.

In order to make the Wigner functions as useful as possible, we choose the maps $\Omega$ so as to maximise the rank of $\Omega^\dagger \Omega$ and therefore the number of states that will be represented.

## A. One spin

The 1-spin Hilbert space is 2-dimensional. The eigenvalues of $S^2, S_3$ are summarised in the following table:

| $S^2$ | $S_3$ | eigenspace dimension |
|---|---|---|
| 3/4 | 1/2 | 1 |
| 3/4 | -1/2 | 1 |

We will use these as a guide for constructing a map $\Omega$ to the 2D harmonic oscillator Hilbert space.

First, we observe that there is a unique state $|\uparrow\rangle$ with eigenvalues $3/4, 1/2$. We will map this to the unique harmonic oscillator state with the same eigenvalues:

$$\Omega |\uparrow\rangle = |10\rangle. \quad (14)$$

As required by condition (ii), this choice preserves norms, since $\langle \uparrow | \uparrow \rangle = \langle 10|10\rangle = 1$. We note that we could have multiplied the right hand side by a phase without spoiling this property. The action of $\Omega$ on the other state is fixed by condition (i) above, via ladder operators:

$$\Omega S_- |\psi\rangle = J_- \Omega |\psi\rangle. \quad (15)$$

This implies that $\Omega |\downarrow\rangle = |01\rangle$. Hence, the map $\Omega$ that we have constructed is uniquely determined by conditions (i) and (ii) above, up to an unimportant overall phase.

Having constructed the map $\Omega$, we compute some Wigner functions. The Wigner function for the state $|\uparrow\rangle$ is

$$W_{10,10}(q_1, q_2, p_1, p_2) = W_{11}(q_1, p_1) W_{00}(q_2, p_2). \quad (16)$$

The Wigner function for the state $|\downarrow\rangle$ is similar, but with the roles of $q_1, p_1$ and $q_2, p_2$ reversed.

The Wigner function (16) is invariant under rotations of the form

$$\begin{pmatrix} q_1 \\ p_1 \\ q_2 \\ p_2 \end{pmatrix} \mapsto \begin{pmatrix} \cos\theta & \sin\theta & 0 & 0 \\ -\sin\theta & \cos\theta & 0 & 0 \\ 0 & 0 & \cos\theta & \sin\theta \\ 0 & 0 & -\sin\theta & \cos\theta \end{pmatrix} \begin{pmatrix} q_1 \\ p_1 \\ q_2 \\ p_2 \end{pmatrix}. \quad (17)$$

It follows that this Wigner function can be written as a function of $x_1, x_2, x_3 \in \mathbb{R}^3$, defined using the Hopf fibration by

$$x_i = \begin{pmatrix} q_1 - \mathrm{i}p_1 & q_2 - \mathrm{i}p_2 \end{pmatrix} \sigma_i \begin{pmatrix} q_1 + \mathrm{i}p_1 \\ q_2 + \mathrm{i}p_2 \end{pmatrix}. \quad (18)$$

Observe that the Wigner function of Eq. (16) is a mapping from $\mathbb{R}^4$ to $\mathbb{R}$. A new function $\mathbb{W}: \mathbb{R}^3 \to \mathbb{R}$ may be constructed, making use of the above coordinate transform, in such a way that Eq. (16) can be rewritten in the form

$$\mathbb{W}(x_1, x_2, x_3) = W_{10,10}(q_1, q_2, p_1, p_2) = -\frac{1}{\pi^2} e^{-r} L_1(x_3+r) \quad (19)$$

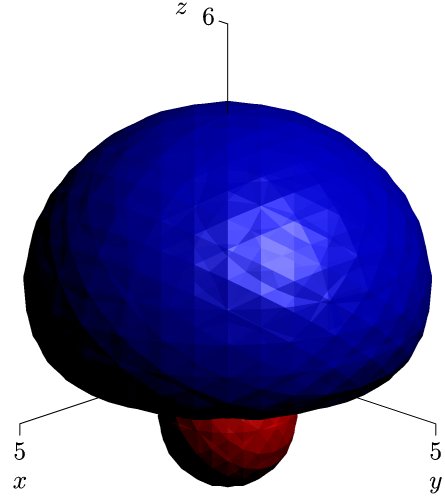

FIG. 1: (Color online) Isosurfaces of $\mathbb{W}$ for the state $|\uparrow\rangle$. This Wigner-like function is positive within the blue regions and negative within the red regions. If viewed in adobe reader these figures are interactive. For the non-interactive figures corresponding to those of the published version of this work please see version one of this preprint.

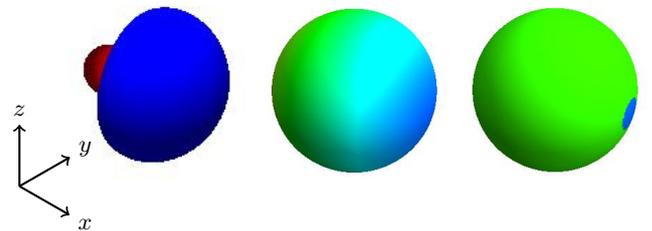

FIG. 2: (Color online) Representation of the 1-spin state $(|\uparrow\rangle + |\downarrow\rangle)/\sqrt{2}$ as a function in $\mathbb{R}^3$ (top), a function on the sphere (middle), and a point on the Bloch sphere (bottom). Blue indicates positive, red indicates negative, and green indicates 0. This figure is not interactive.

(where $r = \sqrt{x_1^2 + x_2^2 + x_3^2}$ and we have used $\mathbb{W}$ instead of $W$ to distinguish the function with transformed arguments form the original Wigner function). Isosurfaces for this function have been plotted in figure 1.

We have also found it useful to represent our Wigner functions as functions on the sphere. Given a Wigner-like function $\mathbb{W}$, we can integrate out the radial degree of freedom and define another Wigner-like function of the

surface of a sphere

$$\mathbb{W}_{\mathbb{S}}(\theta, \phi) = \frac{\pi}{4} \int_0^\infty \mathbb{W}(x, y, z) r \, dr \qquad (20)$$
$$= \frac{\pi}{4} \int_0^\infty \mathbb{W}(r \sin\theta \cos\phi, r \sin\theta \sin\phi, r \cos\theta) r \, dr.$$

Although this may remove some information it does allow for direct comparison with preceding work [4–9]. For this reason we have included, throughout this paper, plots of $\mathbb{W}$ together with those of $\mathbb{W}_{\mathbb{S}}$. Note that the integral here is with respect to $r\,dr$, and not $r^2\,dr$ as one might expect from the volume form of $\mathbb{R}^3$, for the following reason. If the original state $|\psi\rangle$ is normalised so that $\langle\psi|\psi\rangle = 1$ then the Wigner function on the sphere has integral 1:

$$\int_{\mathbb{S}} \mathbb{W}_{\mathbb{S}}(\theta, \phi) \sin\theta \, d\theta \, d\phi = 1. \qquad (21)$$

The representation as functions on the sphere allows us to compare with the Bloch sphere representation: see figure 2.

More generally, it can be shown that the Wigner function $W(q_1, p_1, q_2, p_2)$ for any operator that commutes with $S^2$ can be written as a function $\mathbb{W}$ on $\mathbb{R}^3$, and hence that the Wigner function for any eigenstate of $S^2$ is a function of three variables (see appendix A). On the 1-spin Hilbert space $S^2$ has only one eigenvalue, so all Wigner functions can also be written as functions $\mathbb{W}$ on $\mathbb{R}^3$.

### B. 2 spins

The eigenvalues and eigenspaces of the operators $S_3, S^2$ are:

| $S^2$ | $S_3$ | eigenspace dimension |
|---|---|---|
| 2 | 1 | 1 |
| 2 | 0 | 1 |
| 2 | -1 | 1 |
| 0 | 0 | 1 |

As above, we use these to construct a map $\Omega$ to the 2D harmonic oscillator Hilbert space. First, we identify the state $|\uparrow\uparrow\rangle$ with eigenvalues $(2, 1)$ with the unique harmonic oscillator state with the same angular momentum eigenvalues:

$$\Omega |\uparrow\uparrow\rangle = |20\rangle \qquad (22)$$

The action of $\Omega$ on the other eigenstates of $S^2$ with eigenvalue 2 is fixed by equation (15) above. The state with eigenvalues $(0,0)$ is again identified with a harmonic oscillator state with the same eigenvalues:

$$\Omega \frac{1}{\sqrt{2}} (|\uparrow\downarrow\rangle - |\downarrow\uparrow\rangle) = |00\rangle \qquad (23)$$

This completes the construction of the map $\Omega$. We now consider to what extent the map $\Omega$ is unique. The only choices that we made in specifying $\Omega$ were choices of phase in equations (22), (23). So there is apparently a 2-parameter family of operators $\Omega$ satisfying conditions (i) and (ii). Now one of these parameters corresponds to a choice of overall phase, so is unimportant. The other parameter can be compensated by making a rotation of the form (17), so is again unimportant. Thus the map $\Omega$ is uniquely determined by conditions (i) and (ii).

Wigner functions can now easily be written down. Here are a few examples of Wigner functions for operators that commute with $S^2$. These are all invariant under the rotation (17), so can be written as functions on $\mathbb{R}^3$. Some of these Wigner functions have been plotted in figure 3.

| state | $\mathbb{W}$ |
|---|---|
| $|\uparrow\uparrow\rangle$ | $\frac{1}{\pi^2} e^{-r} L_2(r + x_3)$ |
| $\frac{1}{\sqrt{2}}(|\uparrow\downarrow\rangle + |\downarrow\uparrow\rangle)$ | $\frac{1}{\pi^2} e^{-r} L_1(r + x_3) L_1(r - x_3)$ |
| $|\downarrow\downarrow\rangle$ | $\frac{1}{\pi^2} e^{-r} L_2(r - x_3)$ |
| $\frac{1}{\sqrt{2}}(|\uparrow\downarrow\rangle - |\downarrow\uparrow\rangle)$ | $\frac{1}{\pi^2} e^{-r}$ |

It is worth comparing our construction with those of [4, 5] at this point. Their construction yields interesting Wigner functions for the triplet states (i.e. eigenstates of $S^2$ with eigenvalue 2) but not for the singlet state, which is not defined. In contrast, our construction yields non-zero Wigner functions for all states, both the triplets and the singlet.

For some states the Wigner function $W$ is not invariant under the rotation of Eq. (17). We have not been able to determine a way to reduce such Wigner functions to functions $\mathbb{W}$ of three variables or functions $\mathbb{W}_{\mathbb{S}}$ on the sphere. Indeed, Wigner functions for operators that do not commute with $S^2$ are not invariant under the rotation of Eq. (17) and we do not know if they can be reduced to sensible analogs of $\mathbb{W}$ or $\mathbb{W}_{\mathbb{S}}$. For example, the Wigner function for the operator $|\uparrow\uparrow\rangle (\langle\uparrow\downarrow| - \langle\downarrow\uparrow|)/\sqrt{2}$ is

$$W(q_1, p_1, q_2, p_2) = \frac{\sqrt{2}}{\pi^2} e^{-q_1^2 - p_1^2 - q_2^2 - p_2^2} (q_1 - i p_1)^2. \qquad (24)$$

### C. Three spins

The eigenvalues and eigenspaces of the operators $S^2, S_3$ are:

| $S^2$ | $S_3$ | eigenspace dimension |
|---|---|---|
| 15/4 | 3/2 | 1 |
| 15/4 | 1/2 | 1 |
| 15/4 | -1/2 | 1 |
| 15/4 | -3/2 | 1 |
| 3/4 | 1/2 | 2 |
| 3/4 | -1/2 | 2 |

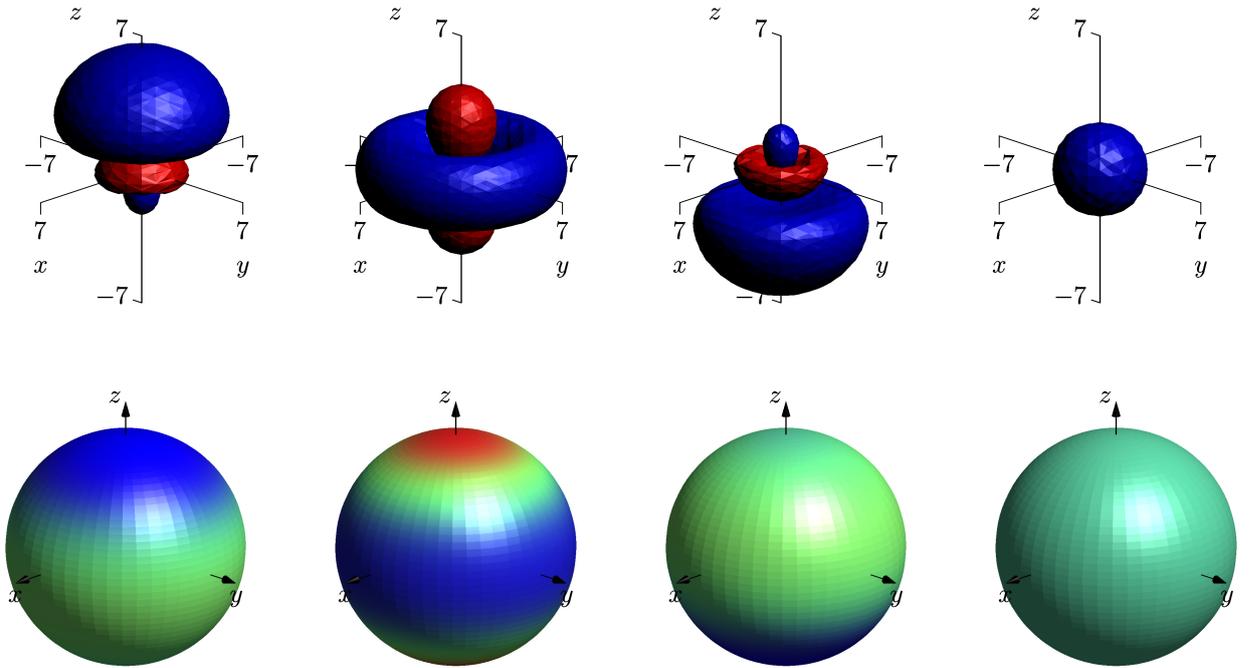

FIG. 3: (Color online) Isosurfaces of Wigner functions for the triplet states $|\uparrow\uparrow\rangle$, $(|\uparrow\downarrow\rangle + |\downarrow\uparrow\rangle)/\sqrt{2}$ and $|\downarrow\downarrow\rangle$, and the singlet state $(|\uparrow\downarrow\rangle - |\downarrow\uparrow\rangle)/\sqrt{2}$. The Wigner functions are positive within the blue regions and negative within the red regions. If viewed in adobe reader these figures are interactive. For the non-interactive figures corresponding to those of the published version of this work please see version one of this preprint.

Note that, unlike in the 1- and 2-spin Hilbert spaces, some of the eigenspace dimensions are greater than 1. This means that our method will be able to construct Wigner functions only for a subspace of the 3-spin Hilbert space.

The construction of Wigner functions proceeds as above. First we fix the action of $\Omega$ on states which are anihilated by $S^+$, guided by the eigenvalues of $S^2$ and $S_3$:

$$\Omega |\uparrow\uparrow\uparrow\rangle = A |30\rangle$$
$$\Omega \frac{1}{\sqrt{3}}\left(|\downarrow\uparrow\uparrow\rangle + e^{2\pi i/3} |\uparrow\downarrow\uparrow\rangle + e^{4\pi i/3} |\uparrow\uparrow\downarrow\rangle\right) = B |10\rangle$$
$$\Omega \frac{1}{\sqrt{3}}\left(|\downarrow\uparrow\uparrow\rangle + e^{-2\pi i/3} |\uparrow\downarrow\uparrow\rangle + e^{-4\pi i/3} |\uparrow\uparrow\downarrow\rangle\right) = C |10\rangle.$$

Here $A, B, C$ are complex parameters. The action of $\Omega$ on the rest of the Hilbert space is fixed by equation (15).

The choices made in the above three equations guarantee that $\Omega$ satisfies condition (i). Condition (ii) is satisfied only if the parameters $A, B, C$ satisfy

$$|A|^2 = 1, \quad |B|^2 + |C|^2 = 1 \qquad (25)$$

These constraints reduce the number of parameters in the choice of $\Omega$ by two, leaving four parameters. One of these parameters corresponds to a choice of overall phase, and another corresponds to rotations of the form (17), so the total number of effective parameters is two.

As in previous examples, the map $\Omega$ can be used to construct Wigner functions; we omit the details.

### D. $N$ spins

As the number of spins increases beyond 3, the number of parameters in the choice of $\Omega$ grows. However, the action of $\Omega$ on the "outer shell" (the eigenspace of $S^2$ with maximal eigenvalue) is always fixed, up to a phase. This is because the eigenspaces of $S_3$ in the outer shell always have dimension 1. The action of $\Omega$ on the outer shell is specified up to a phase by the equation

$$\Omega |\uparrow \ldots \uparrow\rangle = |N0\rangle \qquad (26)$$

and equation (15). Here is a sample of Wigner functions obtained from states in the outer shell (where $|\otimes \uparrow\rangle = |\uparrow \ldots \uparrow\rangle$ and $|\otimes \downarrow\rangle = |\downarrow \ldots \downarrow\rangle$ are spin coherent states [28]):

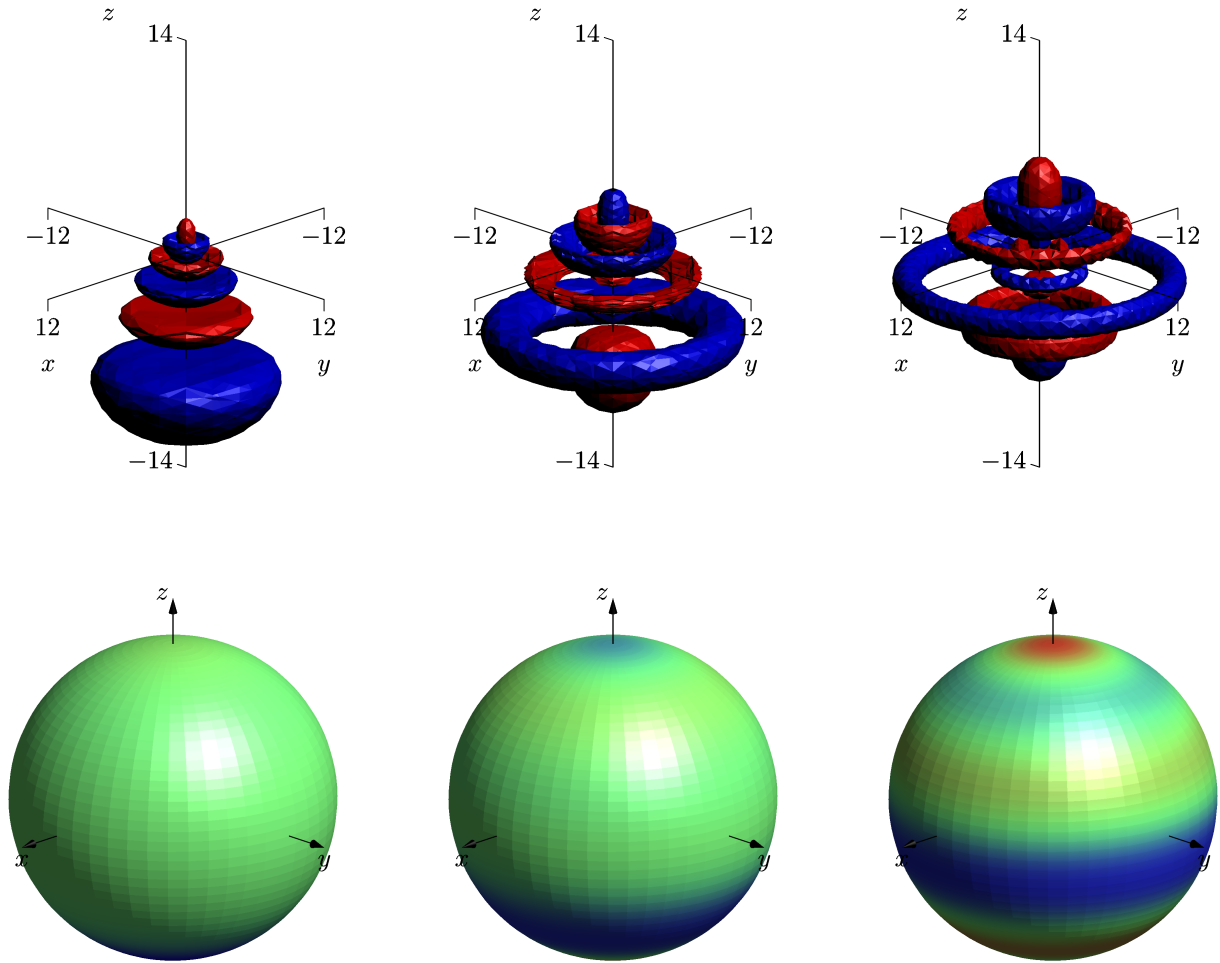

FIG. 4: (Color online) Wigner functions for Fock states with 5 spins represented on $\mathbb{R}^3$ (top row), the sphere (middle row), and the upside-down sphere (bottom row). The precise states plotted are (from left to right) $|\downarrow\downarrow\downarrow\downarrow\downarrow\rangle$, $\sqrt{1/5}S_+|\downarrow\downarrow\downarrow\downarrow\downarrow\rangle$ and $\sqrt{1/40}(S_+)^2|\downarrow\downarrow\downarrow\downarrow\downarrow\rangle$. If viewed in adobe reader these figures are interactive. For the non-interactive figures corresponding to those of the published version of this work please see version one of this preprint.

| Operator | $\mathbb{W}$ |
|---|---|
| $\|\otimes\uparrow\rangle\langle\otimes\uparrow\|$ | $\frac{(-1)^N}{\pi^2}e^{-r}L_N(r+x_3)$ |
| $\|\otimes\downarrow\rangle\langle\otimes\downarrow\|$ | $\frac{(-1)^N}{\pi^2}e^{-r}L_N(r-x_3)$ |
| $\|\otimes\uparrow\rangle\langle\otimes\downarrow\| + \|\otimes\downarrow\rangle\langle\otimes\uparrow\|$ | $\frac{e^{-r}}{\pi^2 N!}\left[(x_1+\mathrm{i}x_2)^N + (x_1-\mathrm{i}x_2)^N\right]$ |

The results we present in Fig. 4 for spin Fock states bear striking resemblance to those found in [4, 5]. In [10] the Wigner function of [4, 5] was used to examine distinct superpositions of spin coherent states (so called spin Schrödinger cats). In Fig. 5 we present $\mathbb{W}$ and $\mathbb{W}_\mathbb{S}$ for two example spin coherent states, the coherent superposition (cat) and their incoherent statistical mixture. In each case we see intuitively reasonable results that are in line with both our expectations and the results given by [4, 5].

Consider $\mathbb{W}$ and $\mathbb{W}_\mathbb{S}$ for the cat and the statistical mixture shown in Fig. 5. Notice that the functions for the coherent superposition of distinct spin coherent states (cat) contain oscillating fringes whilst those of the statistical mixture do not. As with Wigner's original function, the oscillating fringes indicate non-classical characteristics a quantum superposition of states.

We note that, unlike the Wigner function presented in [4, 5], our definition also allows for representation of some (but not all) states below the outer shell (for example, the singlet state for two spins as already discussed). A more detailed investigation is beyond the scope of this paper. We also note that the Wigner-like function $\mathbb{W}$ for spin coherent states has more than one local maximum. As such comparison to the Wigner function for coherent states of the harmonic oscillator must be done with care. However, for large $N$, as with the function presented in [4, 5], $\mathbb{W}_{\mathbb{S}}$ becomes increasingly like the expected Gaussian distribution.

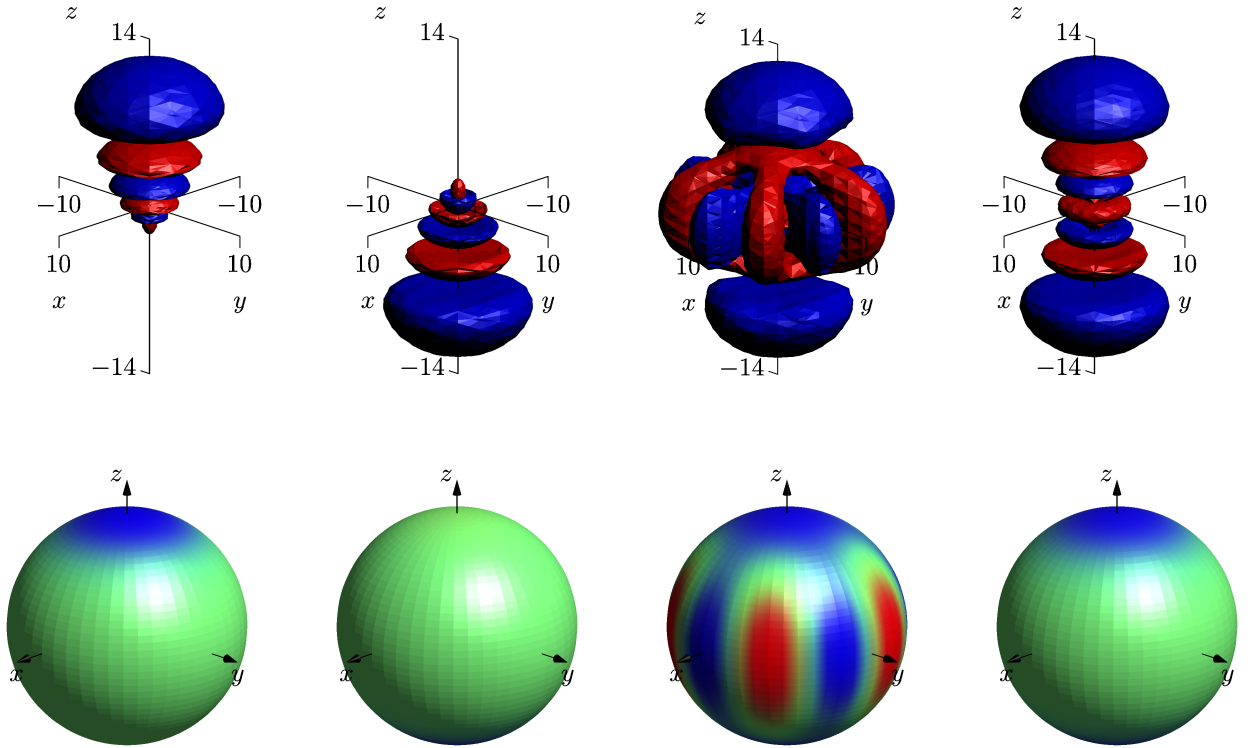

FIG. 5: (Color online) Wigner functions for 5 spins. From left to right: the "up" state $|\uparrow\uparrow\uparrow\uparrow\uparrow\rangle$, the "down" state $|\downarrow\downarrow\downarrow\downarrow\downarrow\rangle$, the "cat" (or GHZ) state $|\uparrow\uparrow\uparrow\uparrow\uparrow\rangle + |\downarrow\downarrow\downarrow\downarrow\downarrow\rangle$ and the statistical sum of the up and down states. The label G indicates a hidden Gaussian-like peak. If viewed in adobe reader these figures are interactive. For the non-interactive figures corresponding to those of the published version of this work please see version one of this preprint.

### E. Squeezed states

As a final example application we consider the spin analogue of squeezed coherent states. Recall that squeezed states for the harmonic oscillator can be defined to be states of the form $\exp(\beta(a^\dagger)^2 - \bar{\beta}a^2)|\alpha\rangle$, with $|\alpha\rangle$ a coherent state and $\beta \in \mathbb{C}$ a parameter. Analogously, we define squeezed states in the system of $N$ spins to be states of the form

$$\exp(\beta(S_+)^2 - \bar{\beta}(S_-)^2)|\alpha; N, l\rangle, \quad (27)$$

with $|\alpha; N, l\rangle$ an spin coherent state [28]. Plots of the associated functions $\mathbb{W}$ and $\mathbb{W}_{\mathbb{S}}$ are shown in figure 6. When $|\beta|$ is small the Wigner-like functions of the squeezed spin states resemble the Wigner functions of squeezed harmonic oscillator states, but for large values of $|\beta|$ the Wigner-like functions spread out and meet themselves on the opposite side of the sphere, leading to interesting interference effects. We note that these results provide a good intuitive and accurate picture of the quantum state of the system.

## VI. CONCLUSIONS

We have presented a new quasiprobability distribution function for ensembles of spins/qubits that has many properties in common with the Wigner function for systems of continuous variables. Our function takes the form of a continuous distribution in $\mathbb{R}^4$ which can, in many cases, be reduced via symmetry arguments to $\mathbb{R}^3$ and the sphere. It enables clear graphical representation of a wide variety of states, including Fock states, spin-coherent states, squeezed states, superpositions and statistical mixtures. Our definition respects the group of rotations generated by $S_1, S_2, S_3$ just as Wigner's original function respects the action of the position and momentum operators. In addition, this property leads to our function exhibiting characteristics with clear parallels to other representations such as the Bloch sphere. More-over, even for relatively large $N$ it is easy to compute both $\mathbb{W}$ and $\mathbb{W}_\mathbb{S}$ using standard mathematical software (to aid calculation of $\mathbb{W}_\mathbb{S}$ analytical expressions are derived in appendix B).

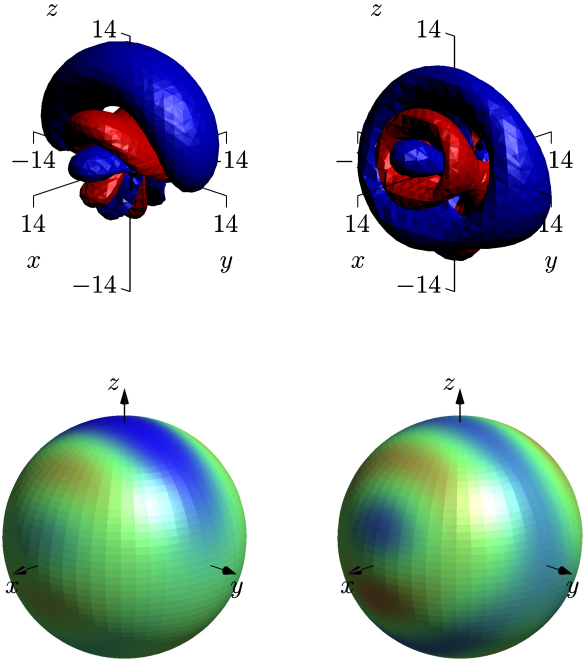

FIG. 6: (Color online) Wigner-like functions for squeezed states $\exp(\beta(S_+)^2 - \bar{\beta}(S_-)^2) |\uparrow\uparrow\uparrow\uparrow\uparrow\rangle$. The left column shows $\beta = 0.1$ and the right shows $\beta = 0.2$. The spherical function $\mathbb{W}_\mathbb{S}$ has been plotted from two opposing viewpoints for greater clarity. If viewed in adobe reader these figures are interactive. For the non-interactive figures corresponding to those of the published version of this work please see version one of this preprint.

In addition to spin Wigner functions with a continuous phase space, a Wigner function can be defined on a discrete phase space [15, 29, 30]. The discrete Wigner functions are characterised best in a prime number dimensions, however can be extended to composite systems. The discrete Wigner function [15] has many very nice mathematical properties but does not lend itself well to graphical interpretation. While the function that we propose in this paper does not share all of the nice mathematical properties of the discrete Wigner function, it is suited to graphical representation, providing intuitive useful information on the nature of the systems quantum state.

We note that, as with continuous systems, it is possible to produce a Husimi Q phase space representation using a projection over spin coherent states of the system's density operator (see, for example, [31–33]). The spin versions of these Husimi Q functions have a similar limitation to those of the continuous case. That is, unlike Wigner functions, the Husimi Q functions do not clearly show interference for macroscopically distinct quantum coherent superpositions of states. For direct comparison between the spin Wigner functions of [4, 5] and the spin Husimi Q functions the we refer the reader to [10] and [33]. We also note that unlike our function the spin Husimi Q function and the spin Wigner functions of [4, 5] are limited to one angular momentum shell.

Our definition allows us to go beyond previous attempts at constructing a continuous Wigner function for ensembles of spins [4–9]. Specifically we include states with different eigenvalues of $S^2$. For example, with two spins our construction yields a non-trivial Wigner function for the singlet state. Nevertheless, not all states can be represented and for three or more spins/qubits information is still lost in passing from harlthe Hilbert space to Wigner functions. The possibility of the existence of a continuous distribution that simultaneously is graphically intuitive and forms a complete representation of the density matrix remains an open question. Nevertheless, we believe that our quasiprobability distribution provides a useful new tool for the analysis and visualisation of quantum states.

### Acknowledgments

MJE is very grateful to W.J. Munro for his hospitality and the generous support of the National Institute of Informatics, Japan where the idea for this work developed through interesting and informative discussions. We also thank J.H. Samson for his input and for suggesting several improvements to the manuscript. We would like to thank the creators of asymptote http://asymptote.sourceforge.net for making their software, which we used to generated the interactive plots in this manuscript, available for general use.

### Appendix A: Rotational invariance of the Wigner function

Our construction of the three-dimensional Wigner-like function $\mathbb{W}$ made essential use of the invariance of the 4-dimensional Wigner function $W$ associated to an eigenstate of $S^2$ under rotations specified in Eq. (17). A short proof of this property follows.

Let $H^{(1)} = \frac{1}{2}(\hat{q}^2 + \hat{p}^2)$ denote the hamiltonian for the 1-dimensional harmonic oscillator and let $\sigma$ denote any operator on the harmonic oscillator Hilbert space. Then the Wigner function for $[iH, \sigma]$ is related to that for $\sigma$ as follows:

$$W_{[iH^{(1)},\sigma]}(q,p) = \left(p\frac{\partial}{\partial q} - q\frac{\partial}{\partial p}\right)W_\sigma(q,p). \qquad (A1)$$

Eq. (A1) may be proved by elementary operations:

$$\frac{1}{\pi}\int_{-\infty}^{\infty}\langle q-y|\left[\tfrac{\mathrm{i}}{2}\hat{q}^2,\sigma\right]|q+y\rangle\, e^{2\mathrm{i}py}$$
$$=\frac{1}{\pi}\int_{-\infty}^{\infty}\tfrac{\mathrm{i}}{2}\left[(q-y)^2-(q+y)^2\right]\langle q-y|\,\sigma\,|q+y\rangle\, e^{2\mathrm{i}py}$$
$$=-q\frac{\partial}{\partial p}\frac{1}{\pi}\int_{-\infty}^{\infty}\langle q-y|\,\sigma\,|q+y\rangle\, e^{2\mathrm{i}py}$$

$$\frac{1}{\pi}\int_{-\infty}^{\infty}\langle q-y|\left[\tfrac{\mathrm{i}}{2}\hat{p}^2,\sigma\right]|q+y\rangle\, e^{2\mathrm{i}py}$$
$$=\frac{1}{\pi}\int_{-\infty}^{\infty}\tfrac{\mathrm{i}}{2}\left[-\left(\frac{\partial^2}{\partial q^2}\langle q-y|\right)\sigma\,|q+y\rangle\right.$$
$$\left.+\langle q-y|\,\sigma\left(\frac{\partial^2}{\partial q^2}|q+y\rangle\right)\right]e^{2\mathrm{i}py}$$
$$=\frac{1}{\pi}\int_{-\infty}^{\infty}\frac{\mathrm{i}}{2}\frac{\partial^2}{\partial q\partial y}\left(\langle q-y|\,\sigma\,|q+y\rangle\right)e^{2\mathrm{i}py}$$
$$=p\frac{\partial}{\partial q}\frac{1}{\pi}\int_{-\infty}^{\infty}\left(\langle q-y|\,\sigma\,|q+y\rangle\right)e^{2\mathrm{i}py}.$$

Suppose now that $|\psi\rangle$ is an eigenstate of $S^2$, or, equivalently, that the density matrix $\rho=|\psi\rangle\langle\psi|$ commutes with $S^2$. Then $\Omega\rho\Omega^\dagger$ commutes with $J^2$. Now the hamiltonian $H=\hat{q}_1^2+\hat{q}_2^2+\hat{p}_1^2+\hat{p}_2^2$ for the 2-dimensional harmonic oscillator is related to $J^2$ via the equation $J^2=(H^2-1)/4$. Since $H$ has only positive eigenvalues, it follows that $\Omega\rho\Omega^\dagger$ commutes with $H$. Then the 2-dimensional analog of Eq. (A1) implies that

$$\left(\sum_{a=1}^{2} p_a\frac{\partial}{\partial q_a}-q_a\frac{\partial}{\partial p_a}\right)W_\rho(q_1,p_1,q_2,p_2)=0. \quad (A2)$$

Eq. (17) follows from this equation by an integration.

### Appendix B: Analytical expressions for $\mathbb{W}_\mathbb{S}$

When computing the spherical function $\mathbb{W}_\mathbb{S}$ one encounters integrals involving products of Laguerre polynomials and the exponential function. It will be shown below that these integrals can always be evaluated analytically in terms of hypergeometric functions. Employing these formulae results in a significant reduction in the time required to compute the Wigner function numerically.

The spherical Wigner function associated to the operator $|lm\rangle\langle lm'|$ is

$$\mathbb{W}_\mathbb{S} = \frac{\pi}{4}\int_0^\infty W_{l+m,l+m'}(q_1,p_1)W_{l-m,l-m'}(q_2,p_2)r\,\mathrm{d}r$$
$$= \begin{cases} \frac{(-1)^{2l}}{4\pi}\sqrt{\frac{(l+m)!(l-m')!}{(l+m')!(l-m)!}}\left(-\sin\theta e^{\mathrm{i}\phi}\right)^{m'-m} I_{l-m',l+m}^{m'-m}(\cos(\theta)) & m\leq m' \\ \frac{(-1)^{2l}}{4\pi}\sqrt{\frac{(l+m')!(l-m)!}{(l+m)!(l-m')!}}\left(-\sin\theta e^{-\mathrm{i}\phi}\right)^{m-m'} I_{l-m,l+m'}^{m-m'}(\cos(\theta)) & m\geq m', \end{cases} \quad (B1)$$

where we have introduced

$$I_{ij}^\alpha(c) := \int_0^\infty e^{-r}r^{1+\alpha}L_j^\alpha((1+c)r)L_i^\alpha(1-c)r)\,\mathrm{d}r. \quad (B2)$$

An integration by parts results in the equation

$$I_{ij}^\alpha(c) = (1+\alpha+i+j)\int_0^\infty e^{-r}r^\alpha L_j(1+c)r)L_i(1-c)r\,\mathrm{d}r$$
$$-(j+\alpha)\int_0^\infty e^{-r}r^\alpha L_{j-1}^\alpha((1+c)r)L_i^\alpha((1-c)r)\,\mathrm{d}r$$
$$-(i+\alpha)\int_0^\infty e^{-r}r^\alpha L_j^\alpha((1+c)r)L_{i-1}^\alpha((1-c)r)\,\mathrm{d}r. \quad (B3)$$

Each of the integrals on the right of this equation evaluates to a hypergeometric function [27]. The resulting expression simplifies on application of Gauss' relations for contiguous functions to

$$I_{ij}^\alpha(c) = (-1)^j \frac{(i+j+\alpha)!}{i!j!}F(-i,-j;-i-j-\alpha;c^{-2})$$
$$\times\left[(i+j+\alpha+1)c^{i+j}+(j-i)c^{i+j-1}\right]. \quad (B4)$$

While concise, this formula is numerically problematic since the hypergeometric function diverges at $c=0$. Slight rearrangement yields the more practical formulae,

$$I_{ij}^\alpha(c) = \begin{cases} (-1)^{j-i}\frac{(j+\alpha)!}{i!(j-i)!}F(-i,1+j+\alpha;1+j-i;c^2)\left[(i+j+\alpha+1)c^{j-i}+(j-i)c^{j-i-1}\right] & i\leq j \\ \frac{(i+\alpha)!}{j!(i-j)!}F(-j,1+i+\alpha;1+i-j;c^2)\left[(i+j+\alpha+1)c^{i-j}+(j-i)c^{i-j-1}\right] & i\geq j. \end{cases} \quad (B5)$$

[1] E. Wigner, Phys. Rev. **40**, 749 (1932), URL http://dx.doi.org/10.1103/PhysRev.40.749.
[2] S. Deléglise, I. Dotsenko, C. Sayrin, J. Bernu, M. Brune, J.-M. Raimond, and S. Haroche, Nature **455** (2008).
[3] M. F. Riedel, P. Böhi, Y. Li, T. W. Hänsch, A. Sinatra, and P. Treutlein, Nature **464**, 1170 (2010), URL http://dx.doi.org/10.1038/nature08988.
[4] J. Dowling, G. Agarwal, and W. Schleich, Phys Rev A **49**, 4101 (1994).
[5] G. Agarwal, Phys Rev A **24**, 2889 (1981).
[6] J. C. Várilly and J. Gracia-Bondía, Ann. Phys. **190**, 107 (1989).
[7] S. Chumakov, A. Frank, and K. Wolf, Phys Rev A **60**, 1817 (1999), URL http://pra.aps.org/abstract/PRA/v60/i3/p1817_1.
[8] F. T. Arecchi, E. Courtens, R. Gilmore, and H. Thomas, Phys. Rev. A **6**, 2211 (1972), URL http://link.aps.org/doi/10.1103/PhysRevA.6.2211.
[9] J. G.-B. Joseph C Várilly, Annals of Physics **190**, 107 (1989).
[10] M. Everitt, W. Munro, and T. Spiller, Phys. Rev. A **85**, 022113 (2012), URL http://pra.aps.org/abstract/PRA/v85/i2/e022113.
[11] C. Brif and A. Mann, Journal of Physics A: Mathematical and General **31**, L9 (1998), URL http://iopscience.iop.org/0305-4470/31/1/002.
[12] W.-M. Zhang, D. H. Feng, and R. Gilmore, Rev. Mod. Phys. **62**, 867 (1990).
[13] M. Hillery, R. O'Connell, M. Scully, and E. Wigner, Physics Reports **106**, 121 (1984), ISSN 0370-1573.
[14] U. Leonhardt, Phys. Rev. Lett. **74**, 4101 (1995).
[15] W. Wootters, Ann. Phys. **176**, 1 (1987).
[16] D. Kaplan and G. Summerfield, Phys. Rev. **187**, 639 (1969).
[17] J. H. Samson, Journal of Physics A: Mathematical and General **36**, 10637 (2003), URL http://www.iop.org/EJ/abstract/0305-4470/36/42/015.
[18] T. Tilma and K. Nemoto, Journal of Physics A: Mathematical and Theoretical **45**, 015302 (2012), URL http://stacks.iop.org/1751-8121/45/i=1/a=015302.
[19] A. Luis, Phys. Rev. **69** (2004), URL http://dx.doi.org/10.1103/PhysRevA.69.052112.
[20] J. E. Moyal, Mathematical Proceedings of the Cambridge Philosophical Society **45**, 99 (1949).
[21] H. Groenewold, Physica **12**, 405 (1946), ISSN 0031-8914, URL http://www.sciencedirect.com/science/article/pii/S0031891446800594.
[22] W. P. Schleich, *Quantum Optics in Phase Space* (Wiley-VCH, 2001).
[23] H. Hopf, Mathematische Annalen **104**, 637 (1931), ISSN 0025-5831, 10.1007/BF01457962, URL http://dx.doi.org/10.1007/BF01457962.
[24] H. Urbantke, Journal of Geometry and Physics **46**, 125 (2003), ISSN 0393-0440, URL http://www.sciencedirect.com/science/article/pii/S0393044002001213.
[25] P. Jordan, Zeitschr. für Physik **94**, 531 (1935).
[26] J. Schwinger, Technical Report NYO-3071, DOE (1952).
[27] I. Gradshteyn and I. Ryzhik, *Table of integrals, series, and products* (Academic Press, 1965).
[28] J. M. Radcliffe, Journal of Physics A: General Physics **4**, 313 (1971), URL http://stacks.iop.org/0022-3689/4/i=3/a=009.
[29] K. S. Gibbons, M. J. Hoffman, and W. K. Wootters, Phys. Rev. A **70**, 062101 (2004), URL http://link.aps.org/doi/10.1103/PhysRevA.70.062101.
[30] C. Ferrie, Reports on Progress in Physics **74**, 116001 (2011), URL http://stacks.iop.org/0034-4885/74/i=11/a=116001.
[31] S. Chaudhury, A. Smith, B. E. Anderson, S. Ghose, and P. S. Jessen, Nature **461**, 768 (2009).
[32] G. R. Jin, S. Luo, Y. C. Liu, H. Jing, and W. M. Liu, J. Opt. Soc. Am. B **27**, A105 (2010), URL http://josab.osa.org/abstract.cfm?URI=josab-27-6-A105.
[33] C. E. A. Jarvis, D. A. Rodrigues, B. L. Györffy, T. P. Spiller, A. J. Short, and J. F. Annett, J. Opt. Soc. Am. B **27**, A164 (2010), URL http://josab.osa.org/abstract.cfm?URI=josab-27-6-A164.
[34] We note that we will interchangeably use the notation $x \leftrightarrow 1$, $y \leftrightarrow 2$ and $z \leftrightarrow 3$. Our rational for doing this is that many or our mathematical arguments are more easily followed using the numerical notion but in some circumstances we wish to emphasise geometrical properties.